# Modélisation à phase résolue de la transformation des vagues en zone de déferlement sur des fonds rugueux idéalisés

Emile Guélard Ancilotti[1], Damien Sous[1,4], Denis Morichon[1], Patrick Marsaleix[2], Héloïse Michaud[3], Solène Dealbera[1]

1. Université de Pau et des Pays de l'Adour, E2S UPPA, SIAME, Anglet, France.
   *emile.guelard-ancilotti@univ-pau.fr, denis.morichon@univ-pau.fr*
2. CNRS, LEGOS, Toulouse, France.
   *patrick.marsaleix@legos.obs-mip.fr*
3. Shom, Toulouse, France.
   *heloise.michaud@shom.fr*
4. Université de Toulon, Aix Marseille Université, CNRS, IRD, MIO, La Garde, France.
   *damien.sous@mio.osupytheas.fr*

**Résumé :**

La compréhension et la prévision des processus de submersion dans les environnements rocheux nécessitent de prendre en compte la rugosité du fond, qui joue un rôle clé dans les processus de transformation des vagues et dans la dynamique littorale générale. Le présent travail vise à mettre en œuvre une paramétrisation de la dissipation induite par la rugosité dans les modèles de vagues 3D non hydrostatiques à phase résolue, basé sur le code Symphonie (MARSALEIX *et al*., 2019). Le modèle modifié est confronté à des expériences de laboratoire réalisées sur une zone de surf à pente linéaire (DEALBERA *et al*., 2024). Différentes vagues irrégulières y ont été testées sur diverses configurations de rugosité du fond (représentées par des configurations en blocs de tailles et de distributions différentes). La génération de vagues et la dissipation induite par le déferlement ont d'abord été paramétrées par rapport aux données de laboratoire sur fond lisse. Les différents cas de rugosité ont ensuite été étudiés sur la base de deux stratégies distinctes pour la paramétrisation du frottement du fond, à savoir l'approche par la contrainte de fond et l'approche par la traînée volumique de canopée. La performance de ces deux approches est ainsi évaluée en comparant les résultats du modèle et les mesures du profil cross-shore de la hauteur significative des vagues pour différentes configurations de fond. Sur la base de ce travail, des recommandations sur le choix des paramétrisations de la dissipation du fond seront prescrites pour chaque environnement.

**Mots clés :**

Vagues, dissipation, rugosité de fond, friction, traînée, canopée, modèle à phase résolue

# 1. Introduction

La dissipation de la houle sur des fonds marins rugueux/végétalisés suscite un intérêt croissant pour les chercheurs et les acteurs des domaines publics et privés du littoral, en raison de son importance pour les risques littoraux, les activités et les installations côtières. La compréhension détaillée de l'impact des rugosités de fonds sur la propagation de la houle constitue un défi en raison de la complexité des phénomènes hydrodynamiques associés et du manque de littérature, en particulier concernant les fonds rocheux. La modélisation numérique à phase résolue offre une approche précise et efficace pour étudier les processus hydrodynamiques côtiers et représenter l'évolution de la surface libre de façon résolue dans le temps et l'espace. En particulier les modèles 3D (ou pseudo-3D) non hydrostatiques, tel que SYMPHONIE NH (MARSALEIX *et al.*, 2019), capables de simuler à la fois les vagues, les courants à plus basse fréquence et leurs interactions en prenant compte la pression non hydrostatique. Néanmoins, représenter avec précision la rugosité et ses effets dans les modèles numériques reste un défi car très coûteux numériquement et nécessitant des données topo-bathymétriques haute résolution. Deux approches existent à ce jour pour modéliser la dissipation induite par la rugosité sans la représenter explicitement : l'approche par friction sur le fond (Bottom Stress Approach - BSA) et l'approche par traînée volumique de canopée (Canopy Drag Approach - CDA). Classiquement utilisée pour des fonds sableux dans la grande majorité des modèles de vagues, la BSA est utilisée dans SYMPHONIE NH pour représenter la friction en limite de fond et se communiquant au reste de l'écoulement. Néanmoins, des adaptations pour les fonds rocheux avec la CDA sont nécessaires pour améliorer le modèle et reproduire correctement l'hydrodynamique et plus particulièrement la dissipation des vagues dans ces environnements où les rugosités de fond peuvent occuper un volume conséquent dans l'écoulement semblable à une canopée. L'objectif de cette étude est ainsi d'améliorer la compréhension de la propagation de la houle sur fond rugueux en explorant les approches BSA et CDA dans SYMPHONIE NH, pour différents cas de rugosité en conditions idéalisées de laboratoire. Dans un premier temps, la méthodologie associée à l'expérience en canal LEGOLAS et à l'implémentation des paramétrisations optimales pour les approches susmentionnées dans le modèle sera présentée. Puis, les résultats des simulations avec les deux approches seront confrontés aux mesures expérimentales. Enfin, une discussion autour des deux approches suivra dans l'objectif d'améliorer la représentation en modélisation à phase résolue de la propagation des vagues dans ces environnements.

# 2. Méthode

## 2.1. Expérimentation en canal à houle

Les tests ont été réalisés dans le canal CASH (figure 1), d'une longueur de 6 m, d'une largeur de 0.5 m et d'une profondeur de 0.22 m, comportant une pente linéaire de 1/20. Un batteur à houle génère des vagues sur la partie horizontale en amont de la pente. Trois forçages de 5 minutes de vagues irrégulières ont été testés (figure 1), en faisant varier la hauteur significative (Hs : 0.047, 0.061 et 0.071 m) à partir d'un spectre de JONSWAP. Une configuration avec une pente lisse a été utilisée comme référence sans rugosité. Parmi les autres types de fond rugueux testés expérimentalement, nous avons retenu les configurations en quinconce QC2, QC4 et QC6 (figure 1) de dimension croissante (0.01, 0.03 et 0.05 m d'arête). Pour mesurer l'élévation de la surface libre, des sondes résistives ont été déployées dans le canal. Trois alignements de sondes ont été disposés, dont un central avec 20 sondes réparties tous les 17 cm environ dans l'axe du canal. Les profils de Hs ont été obtenus en intégrant les spectres sur la bande de fréquences associée aux vagues gravitaires (0.5*fp* < *f* < 3*fp*, où *fp* est la fréquence pic).

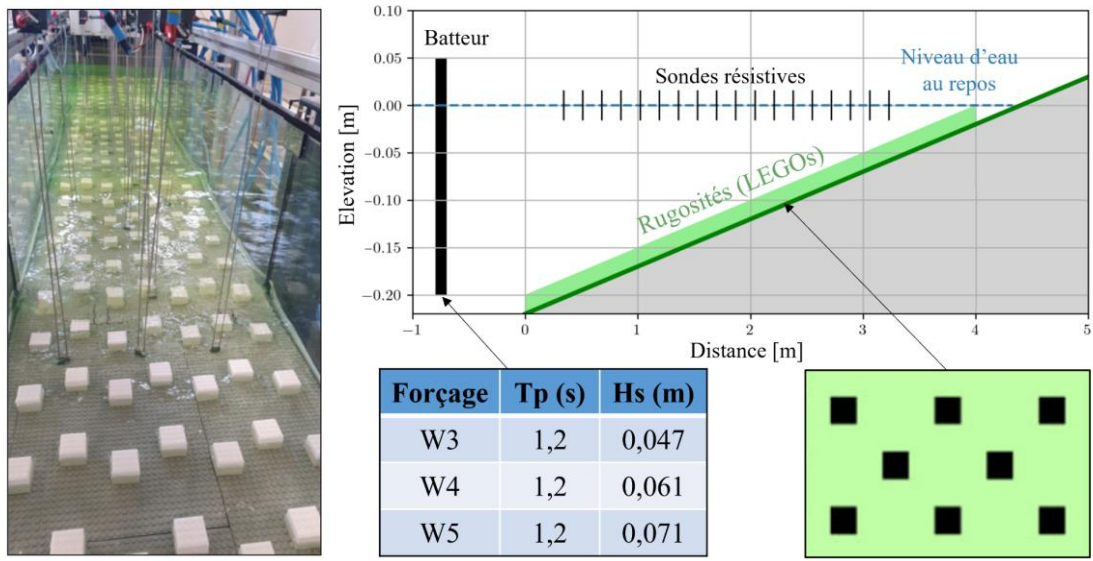

*Figure 1. Photo et schéma de l'expérimentation LEGOLAS accompagné des différents forçages générés au niveau du batteur et d'un exemple de motif de blocs*

2.2. Simulation numérique

Le modèle non-hydrostatique à phase résolue SYMPHONIE NH (MARSALEIX *et al.*, 2019) a été employé pour reproduire l'expérimentation numériquement selon BSA ou CDA. Les forces de résistance $F_{BSA}$ et $F_{CDA}$ comparées dans cette étude s'expriment comme termes sources au sein de l'équation de conservation de quantité de mouvement. Selon BSA, la dissipation est uniquement appliquée au niveau de la couche de fond du modèle (d'une épaisseur $h_{k_{min}}$) via un coefficient de friction $C_f$, fonction du profil logarithmique de la hauteur de rugosité $z_0$ (équation 1 et figure 2). Elle se propage aux autres couches via un coefficient de diffusion $K_M$ inspiré du schéma de fermeture de turbulence de CRAIG & BANNER (1994) pour une stratification neutre (équation 2).

$$F_{BSA}(k_{min}) = C_f u_{k_{min}}|u_{k_{min}}| = (\frac{0.4}{log(\frac{h_{k_{min}}}{z_0})})^2 u_{k_{min}}|u_{k_{min}}| \tag{1}$$

$$F_{BSA}(k) = \frac{\partial}{\partial z_k}(K_M \frac{\partial u_k}{\partial z_k}) \tag{2}$$

Selon CDA, la dissipation est quant à elle appliquée à tous les niveaux de grille allant du fond jusqu'au sommet de la hauteur d'influence de la rugosité $h_R$ (figure 2). $F_{CDA}$ s'exprime alors au travers d'un fonction $f$ de distribution verticale et d'un coefficient de traînée empirique $C_0$ (équation 3), inspiré du coefficient de traînée classique $C_D$ et optimisé selon le forçage de vague et la configuration de fond. $f$ dépend de $h_R$, égale à 4 fois l'écart-type des variations de l'élévation du fond $z_b$ (figure 2), qui représente la hauteur sur laquelle les rugosités sont le plus réparties d'après la fonction de densité de probabilité de $z_b$, décrite par CHUNG et al. (2021).

$$F_{CDA}(k) = C_D u_k|u_k| = fC_0 u_k|u_k| \quad avec\ f = \{\frac{1}{2}[1 - tanh(\frac{h+z(k)+h_R}{0.1h_R})]\} \tag{3}$$

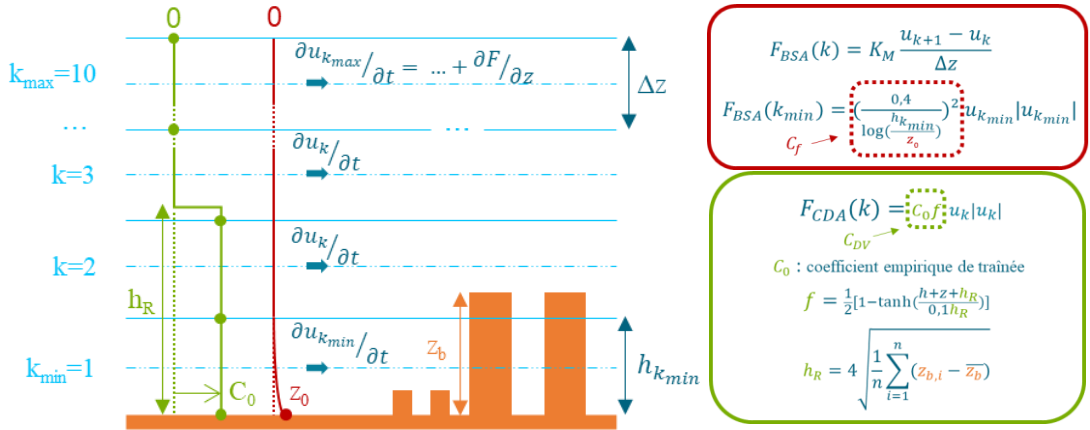

*Figure 2. Représentation schématique de la discrétisation de la colonne d'eau pour calculer la friction sur le fond (rouge) ou la traînée volumique de canopée (vert)*

Pour comparer les deux approches, une première série de simulations a été effectuée selon l'approche BSA, c'est-à-dire en ne prenant pas en compte la dissipation $F_{CDA}$ (= 0), pour trouver les valeurs de $z_0$ optimales (équation 1) en minimisant les erreurs avec les mesures en canal. Une seconde série de simulations a été effectuée selon l'approche CDA en réduisant $z_0$ à $10^{-5}$ m pour négliger la friction en limite de fond et prendre en compte la dissipation multicouche $F_{CDA}$ (active pour les couches de fond comprises dans $h_R$), pour trouver les valeurs de $C_0$ optimales (équation 3).

La configuration des simulations se base sur un canal numérique prolongé deux mètres en amont pour stabiliser la condition d'entrée avec une résolution horizontale de 0.03 m (267 points) et 10 couches sigma sur la verticale. Le pas de temps est fixé à 0.002 s. Les forçages de vagues générés en frontière du domaine numérique sont issus des spectres des séries temporelles d'élévation de la surface libre enregistrées à la première sonde en début de rampe. Ces spectres ont permis de reproduire des séries temporelles synthétiques de surface libre à phase aléatoire, générées 3 m en amont de la rampe pour permettre aux vagues de se mettre en place plutôt que de les générer juste devant la rampe comme dans l'expérimentation. Les paramètres de déferlement du modèle (type critère de Miche, MARSALEIX *et al.*, 2019) ont été calibrés sur les mesures effectuées sur fond lisse selon chaque forçage, pour pouvoir être appliqués aux autres cas rugueux et négliger la dissipation par déferlement. La contribution de la rugosité à la dissipation des vagues est ensuite analysée au travers de l'étude des profils de Hs (tracés à partir des sorties du modèle suivant la même procédure que les expériences) et de l'erreur moyenne quadratique (RMSE, minimale à 0) et l'indice de Willmott (WI, minimale à 1) associés.

### 3. Résultats

Les profils de Hs optimaux obtenus selon chaque approche, forçage et configuration de fond étudiées sont regroupés dans la figure 3, et leurs coefficients optimaux ($z_0$ ou $C_0$) et leurs erreurs associées (RMSE et WI) dans le tableau 1. Les profils de Hs issus de SYMPHONIE NH (représentés par les traits pleins) montrent des valeurs trop élevées dans les faibles profondeurs (à partir de 2 m de distance environ sur la rampe en noir) vis-à-vis des valeurs mesurées en canal (représentées par les croix) selon l'approche BSA. Avec l'ajout de la dissipation additionnelle appliquées aux couches traversées par $h_R$, l'approche CDA montre des profils de Hs plus proches des mesures en canal, notamment dans les faibles profondeurs, associés également à des erreurs plus faibles.

*Tableau 1. Coefficients optimaux et erreurs associées des profils de Hs modélisés selon chaque approche, forçage et configuration de fond*

| Forçage | Fond | BSA | | | CDA | | | |
|---|---|---|---|---|---|---|---|---|
| | | $z_0$ | RMSE | WI | $h_R$ | $C_0$ | RMSE | WI |
| W3 | QC2 | 0.005 | 0.0013 | 0.993 | 0.020 | 15 | 0.013 | 0.993 |
| W3 | QC4 | 0.300 | 0.0020 | 0.983 | 0.044 | 15 | 0.018 | 0.988 |
| W3 | QC6 | 0.300 | 0.0047 | 0.922 | 0.064 | 15 | 0.033 | 0.967 |
| W4 | QC2 | 0.003 | 0.016 | 0.995 | 0.020 | 1 | 0.017 | 0.994 |
| W4 | QC4 | 0.010 | 0.025 | 0.988 | 0.044 | 10 | 0.024 | 0.989 |
| W4 | QC6 | 0.300 | 0.056 | 0.945 | 0.064 | 15 | 0.041 | 0.972 |
| W5 | QC2 | 0.005 | 0.032 | 0.986 | 0.020 | 0.1 | 0.020 | 0.995 |

| | | | | | | | |
|---|---|---|---|---|---|---|---|
| W5 | QC4 | 0.100 | 0.043 | 0.975 | 0.044 | 15 | 0.030 | 0.989 |
| W5 | QC6 | 0.300 | 0.077 | 0.987 | 0.064 | 20 | 0.047 | 0.977 |

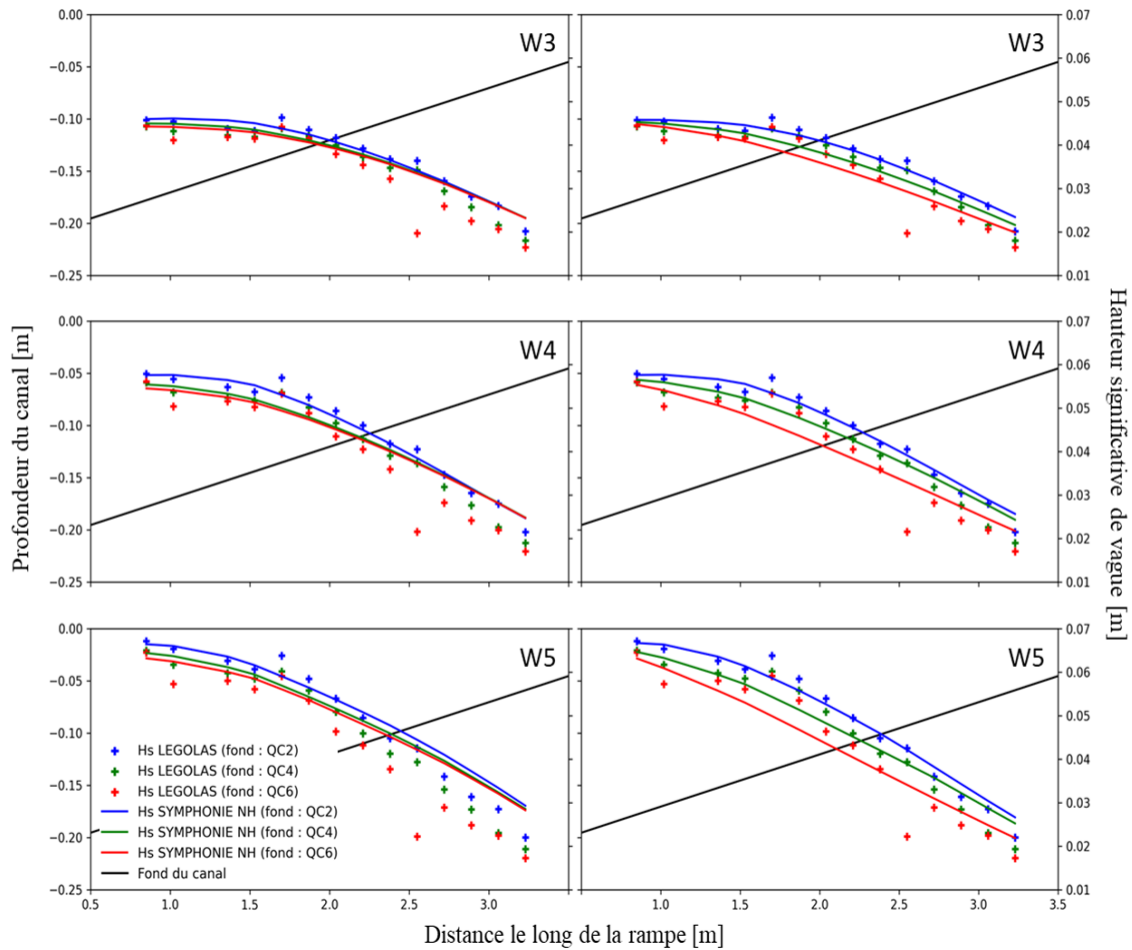

*Figure 3. Profils de Hs mesurés (croix colorées) et modélisés (traits colorés) pour les 3 forçages de vagues W3, W4 et W5, se propageant le long de la rampe (trait noir), selon la configuration de fond QC2, QC4 et QC6 (bleu, vert et rouge) et selon une approche par friction sur le fond (BSA) ou par traînée volumique de canopée (CDA)*

## 4. Discussion

Cette étude vise à approfondir la compréhension de la modélisation des vagues sur fonds rugueux en examinant les méthodes BSA et CDA au sein de SYMPHONIE NH, basé sur l'expérimentation LEGOLAS.

De manière générale, les profils de Hs obtenus montrent des valeurs de Hs trop importantes dans les faibles profondeurs vis-à-vis des mesures en canal, témoignant d'une dissipation sous-estimée par le modèle.

En s'appuyant sur les profils et la moyenne des valeurs de RMSE et WI, la CDA montre une amélioration des résultats, avec des Hs plus proches des mesures en canal avec des valeurs respectives de 0.027 m et 0.985 (contre 0.029 m et 0.975 pour la BSA).

Concernant BSA, cela peut être dû à un problème physique avec une influence des rugosités qui va dépasser la couche de fond du modèle à laquelle la dissipation $F_{BSA}$ est appliquée. De la même façon que cela peut être dû à un problème numérique du fait que le modèle utilise des couches sigma (conservant le même nombre de couches verticale entre l'élévation instantanée de la surface et du fond, soit 10 dans notre cas) qui, du fait de la rampe, vont se trouver très resserrées dans les faibles profondeurs, avec une rugosité hydraulique $z_0$ allant forcément dépasser la couche de fond réduite du modèle en se rapprochant du rivage. Ainsi, bien que plusieurs couches soient censées être affectées par la rugosité, seule la couche de fond du modèle est impactée et la dissipation des couches supérieures n'est pas pleinement prise en compte, sous-estimant ainsi la dissipation totale au niveau des faibles profondeurs.

Selon CDA, en ajoutant la dissipation additionnelle aux couches traversées par $h_R$, les profils de Hs montrent des valeurs modélisées qui se rapprochent des mesures en canal au niveau des faibles profondeurs. Néanmoins, avec cette approche, le modèle semble surestimer la dissipation dans les profondeurs intermédiaires (entre 1 et 2,4 m de distance de la rampe environ) et toujours sous-estimer légèrement la dissipation dans les faibles profondeurs (au-delà de 2,4 m de distance).

Les coefficients $C_0$ trouvés sont pour la plupart de l'ordre de 10 à 20 (ce qui vaut 10 fois plus que les valeurs habituelles de $C_D$ trouvées dans la littérature), sauf pour le forçage le plus faible W3. Ainsi, il apparaît qu'une autre grandeur liée à l'hydrodynamique (e.g., nombre d'Iribarren), la topographie du fond (e.g., densité des rugosités ou solidité frontale) ou les deux (e.g., nombre de Reynolds ou de Keulegan-Carpenter) devrait être prise en compte (en tant que coefficient) pour rendre compte de cette dissipation en fonction de la profondeur (YIN *et al*., 2024).

De plus, seule la part turbulente (quadratique) de $F_{CDA}$ à été prise en compte sans s'intéresser à l'influence de la part laminaire (linéaire), souvent négligeable pour ce type d'écoulement, ou inertielle (associée à la masse ajoutée), qui devrait avoir une influence notable en présence de rugosités conséquentes comme des rochers (VAN GENT, 1993).

**5. Conclusion**

Cette étude sur la propagation des vagues sur des fonds marins rugueux selon l'approche BSA ou CDA offre des perspectives significatives pour comprendre les phénomènes hydrodynamiques côtiers à l'aide de modèles non-hydrostatiques à phase résolue tels que SYMPHONIE NH. L'approche par friction sur le fond (BSA) sous-estime la dissipation des vagues dans les faibles profondeurs car seule la couche de fond est impactée bien que les rugosités la dépassent. Les limitations numériques, telles que l'utilisation de couches sigma, contribuent également à cette sous-estimation dans les faibles profondeurs où cette couche de fond est d'autant plus faible. D'autre part, l'approche par traînée volumique de canopée (CDA) améliore la modélisation de la dissipation dans les faibles profondeurs en impactant toutes les couches de fonds traversées par $h_R$, mais tend à la surestimer dans les profondeurs intermédiaires. Les coefficients $C_0$, bien que supérieurs aux valeurs habituelles de $C_D$, suggèrent la nécessité de considérer d'autres paramètres liés à l'hydrodynamique et à la topographie du fond. L'exploration de la part laminaire et inertielle de l'écoulement doit également être réalisée, en particulier en présence de rugosités importantes comme des rochers. Ainsi, les développements apportés au modèle SYMPHONIE NH pourront ensuite être validés et appliqués dans des conditions réalistes à l'aide de campagnes de mesures sur fonds rocheux (plage sablo-rocheuse d'Erromardie, plate-forme rocheuse d'Ars-en-Ré). Puis, ils pourront être adaptés à d'autres environnements rugueux tels que les fonds récifaux, coralliens ou végétalisés pour enrichir la connaissance sur l'hydrodynamique induite par ces fonds complexes rugueux.

## 6. Remerciements